\documentclass{article}

\usepackage{amsmath,amssymb,amsthm}
\usepackage{color}

\theoremstyle{plain}
\newtheorem{thm}{Theorem}[section]

\theoremstyle{definition}

\newtheorem{ex}[thm]{Example}

\theoremstyle{remark}

\newcommand{\EE}{\mathbb E}

\title{NJst and ASTRID are not statistically consistent under a random model of missing data}
\author{John A. Rhodes \and Michael G. Nute \and   Tandy Warnow}

\begin{document}

\maketitle

\begin{abstract}
Species tree estimation from multi-locus datasets is statistically challenging for multiple reasons, including gene tree heterogeneity across the genome due to incomplete lineage sorting (ILS).
Species tree estimation methods have been developed that operate by estimating gene trees and then using those gene trees to estimate the species tree.
Several of these methods (e.g., ASTRAL, ASTRID, and NJst) are provably statistically consistent under the multi-species coalescent (MSC) model, provided that the gene trees are estimated correctly, and there is no missing data.
Recently, Nute {\em et al.} (BMC Genomics 2018) addressed the question of whether these methods remain statistically consistent under random models of taxon deletion, and asserted that they do so. 
Here we provide a counterexample to one of these theorems, and establish that  ASTRID and NJst are not statistically consistent under an {\em i.i.d.} model of taxon deletion. 

\end{abstract}




\section{Introduction}
Species tree estimation from multi-locus datasets is statistically challenging for multiple reasons, including gene tree heterogeneity across the genome due to incomplete lineage sorting (ILS), gene duplication and loss (GDL), and other processes \cite{maddison1997gene}. 
Because species trees are used in many downstream biological analyses, there is much interest in methods that can estimate species trees given conditions with heterogeneity across the 
genome \cite{DEGNAN2009332,Bravo-biorxiv2019}.

Many methods have been developed to enable these estimations, especially for the specific case where gene tree heterogeneity is due to ILS, which is expected to impact species tree estimation whenever there is a rapid radiation  \cite{rosenberg2013discordance}.
In particular, ILS is believed to be present for many gene trees in the avian \cite{Jarvis2014} and plant \cite{Wickett2014,1kp-nature} phylogenies.

Among the methods that have been developed to estimate species trees in the presence of ILS, the ones that have been most widely adopted (in part for computational reasons) operate in two stages: first they estimate the gene trees, and then they estimate the species tree from the gene trees, using summary statistics.
Many of these ``summary methods",  including ASTRAL \cite{astral,astral2,astral3}, 
ASTRID \cite{astrid}, and NJst \cite{njst}, 
are statistically consistent under the multi-species coalescent (MSC) model \cite{kingman1982coalescent}, which means that as the number of true gene trees increases, the error in the estimated species tree will converge to $0$ almost surely. 

Yet this theoretical statement assumes that there is no missing data (i.e., no species missing from any gene tree). 
Hence, the statistical consistency of these methods when species can be missing from gene trees is an important consideration.
This is the question addressed in \cite{nute-missing}, which examined
the statistical consistency of ASTRAL, ASTRID, and NJst, under two random models of missing data.


%
 
 Theorem 11 of \cite{nute-missing} asserts that NJst
  and ASTRID give statistically consistent estimates of the unrooted species tree topology under the $MSC$ combined with the $M_{iid}$ model of missing taxa (i.e., where species are deleted randomly from 
 the gene trees under an {\em i.i.d.} model).
 However, the theorem is incorrect, as we show by providing a counterexample.
 
\section{The counterexample}

NJst and ASTRID both operate by first computing the
``internode distance matrix" (i.e., for each pair of species, the internode distance matrix is the average, across all the gene trees, of the  distance between the two species in the tree). 
Given the internode distance matrix, 
 ASTRID and NJst  then compute a species tree from this matrix using a distance-based method.
The difference between ASTRID and NJst lies  
only in the choice of distance method (neighbor joining \cite{NJ} for NJst, and FastME \cite{lefort2015fastme} for ASTRID). 
However, as long as the internode distance matrix is sufficiently close to {\em additive} for the species tree (which means it is equal to path distances in the species tree for some non-negative edge weights), both methods are guaranteed to return the species tree.

 The proof of Theorem 11 in \cite{nute-missing} attempts to establish that the matrix of expected pairwise distances is additive on the unrooted topology of the species tree, 
 even when data are missing under these random models. 
 From this property, the statistical consistency of ASTRID and NJst is then
easily established.
However, 
the argument that the matrix of expected pairwise distances is additive is not valid, as the following example shows.

\begin{ex}
Consider the balanced ultrametric species tree on six taxa $a,b,c,d,e,f$
$$\sigma=((a:L+1,(b:1,e:1):L):\epsilon,(c:L+1,(d:1,f:1):L):\epsilon),$$
where $\epsilon$ and $L$ are measured in coalescent units.
We will show that when $L=\infty$, $\epsilon=0$, and $p\in (0,1)$ gives the probability of taxon presence under $M_{iid}$, the expected distances under the combined $MSC+M_{iid}$ model satisfies the strict 4-point inequality
for a tree displaying the resolved quartet $ac|bd$.  

If this is established, then since the expected distances are continuous functions of $\sigma$'s branch lengths in the interval $[0,\infty]$, for sufficiently small $\epsilon>0$ and sufficiently large $L<\infty$ the expected distance will still satisfy the strict 4-point inequality to display $ac|bd$. As  $ac|bd$ is not displayed on $\sigma$, this gives examples of binary species trees with finite edge lengths on which the expected distance is not additive.

We henceforth assume $L=\infty$ and $\epsilon=0$. Under the $MSC$ two lineages entering a  population of infinite duration coalesce in that population with probability 1. Thus in any realization of the coalescent process with no missing taxa, our species tree always leads to 4 lineages entering the population ancestral to the root of $\sigma$: two lineages from $a$ and $c$ and two combined lineages from $be$ and $df$. However,  the $M_{iid}$ model of missing taxa may lead to some of these lineages not being present. As a result, we need to consider how 2, 3, or 4 lineages entering the ancestral population affect the distance.

Let $E_k$ denote the expected number, under the MSC, of internal nodes along the path between two sampled individuals, $x,y,$ in the unrooted gene tree formed when $k$ individuals (including $x,y$) are sampled in a single population of infinite duration. Then $E_2=0$ as in that case the gene tree is simply a single edge, while $E_3=1$, as the gene tree is then a 3-leaf star tree. To compute $E_4$, note that the 3 resolved quartet gene trees each have probability $1/3$, but the number of nodes between $x$ and $y$ on these trees is then $1$, $2$, or $2$. Thus $E_4=5/3$.

On an observed gene tree under the combined $MSC+M_{iid}$ model the probabilities of the each of the lineages $a$ and $b$ being present is $p$, and absent is $q=1-p$. For each of the lineages $be$ and $df$, the probability of presence is $1-q^2$ and absence is $q^2$. Thus some of the expected distances between taxa are

\begin{align*}
\EE( \rho(a,b))&=\EE (\rho(a,d))=\EE( \rho(b,c))=\EE(\rho(c,d))\\
&=p^2\big (q\cdot q^2 (E_2+p\cdot 1)+ pq^2(E_3+p\cdot1).\\
& \hskip 1.in+q(1-q^2))(E_3+p\cdot1) +p(1-q^2)(E_4+p\cdot1)\big )\\
&=p^2\left (p+q^3\cdot 0+(q+q^2-2q^3)1+(1-q-q^2+q^3)\frac 53\right )\\
&=p^2\left (p-\frac 13q^3-\frac 23 q^2-\frac 23q +\frac53\right ),\\
\  \\
\EE( \rho(a,c))&=p^2\left (q^2\cdot q^2 E_2+2q^2(1-q^2)E_3 +(1-q^2)(1-q^2)E_4\right )\\
&=p^2\left (q^4\cdot 0+(2q^2-2q^4)1+(1-2q^2+q^4)\frac 53\right )\\
&=p^2\left (-\frac 13q^4-\frac 43 q^2+ \frac53\right ),\\
\end{align*}
\begin{align*}
\EE( \rho(b,d))&=p^2\big (q^2 (E_2+2pq\cdot1+p^2\cdot2)+2pq(E_3+2pq\cdot1+p^2\cdot2)\\
&\hskip1.5in+p^2(E_4+2pq\cdot1+p^2\cdot2)\big )\\
&=p^2\left (2p+q^2\cdot 0+(2q-2q^2)1+(1-2q+q^2)\frac 53\right )\\
&=p^2\left (2p-\frac 13q^2-\frac 43 q+\frac 53\right ).
\end{align*}
One can then verify that
$$
\EE( \rho(a,b)+\EE (\rho(c,d)) =\EE( \rho(a,d))+\EE (\rho(b,c))> \EE( \rho(a,c))+\EE (\rho(b,d))\label{eq:4pt}
$$
 which is the strict 4-point inequality for a tree displaying the quartet $ac|bd$.

\end{ex}

In other words, the average internode distance matrix computed by ASTRID and NJst can, under some conditions, converge to a matrix that is additive but for a tree topology that is different from the true species tree.
As a result, neither ASTRID nor NJst can be statistically consistent under such a condition, since they will return the wrong tree topology.
In other words, ASTRID and NJst will be {\em positively misleading} (a worse property than being statistically inconsistent) for such a model condition.

\section{Conclusions}
This paper shows that ASTRID and NJst, two species tree estimation methods, are not statistically consistent estimators of the species tree under a model  of missing data.
This result may suggest that species tree estimation in the presence of missing data should be restricted to other methods, such as ASTRAL, that do remain statistically consistent when the gene trees are missing species under an {\em i.i.d.} model.
A recent study
 \cite{genefiltering} showed that ASTRID remains reasonably competitive with ASTRAL under missing data models.
However, future work is needed to explore this question more broadly, to understand the impact in practice of missing data.


\end{document}